\documentclass[prl,twocolumn]{revtex4}
\usepackage{graphicx}

\usepackage{epstopdf} \DeclareGraphicsExtensions{.eps, .pdf}

\begin{document}

\def\K{{\bf{K}}}
\def\Q{{\bf{Q}}}
\def\Gbar{\bar{G}}
\def\tk{\tilde{\bf{k}}}
\def\k{{\bf{k}}}
\def\q{{\bf{q}}}

\title{High energy kink in the single particle spectra of the two-dimensional Hubbard model}
\author{Alexandru Macridin$^{1}$,  M.\ Jarrell$^{1}$, Thomas Maier$^{2}$, 
D.\ J.\ Scalapino$^{3}$}
\address{
$^{1}$ University of Cincinnati, Cincinnati, Ohio, 45221, USA\\
$^{2}$ Oak Ridge National Laboratory, Oak Ridge, Tennessee, 37831, USA\\
$^{3}$ University of California, Santa Barbara, CA 93106-9530}

\date{\today}

\begin{abstract}

 Employing dynamical cluster quantum Monte Carlo calculations we show that the   single particle spectral weight $A(k,\omega)$ 
of the one-band two-dimensional   Hubbard model displays a high energy kink in the quasiparticle dispersion   followed by a steep dispersion of a broad peak similar to recent ARPES results   reported for the cuprates.  Based on the agreement between the Monte Carlo  results and a simple calculation which couples the quasiparticle to spin   fluctuations, we conclude that the kink and the broad spectral feature in the   Hubbard model spectra is due to scattering with damped high energy spin   fluctuations.

\end{abstract}

\maketitle


{\em{Introduction.}} Angle-resolved photoemission spectroscopy (ARPES) has revealed much about the cuprates, including the energy scales associated with the $d$-wave gap\cite{z_shen_93} and a low energy kink presumably associated with strong electron-phonon coupling\cite{a_lanzara_01}. Recent ARPES experiments have revealed a high-energy (HE) kink and a waterfall structure\cite{j_graf_06a,w_meevasana,t_valla_06,j_chang_06,xie_06,pan_06}, in which the band dispersion broadens and falls abruptly at binding energies below $\approx 0.35$ eV.  The origin of this kink has been attributed to a crossover from the quasiparticle (QP) to the Mott-Hubbard band\cite{w_meevasana,wang_06} the settlement of spin-charge separation\cite{j_graf_06a}, or interaction of the quasiparticles (QP) with spin fluctuation excitations\cite{grober_00,odashima_05,ronning_05,manousakis_06,t_valla_06}.

In this Letter, we study the single particle spectral weight $A(k,\omega)$ of the one-band 2D Hubbard model with near-neighbor hopping $t$ and Coulomb interaction $U$ in the regime where $U$ is comparable to the bandwidth $W=8t$ and in the doping range relevant for cuprate superconductors.  The single-band Hubbard model is believed to describe the low-energy physics of the cuprates down to energies of $\approx 2t$ below Fermi surface (FS).  Surprisingly, the calculated spectra of the single band model are remarkably similar to the experimental ones down to binding energies of $\approx 4t-5t$.  They reveal a sharp QP feature down to a kink energy $E_{kink}$, followed by a steep dispersion of a broad waterfall structure.  We find that these features are accurately captured by a renormalized second order (RSO) approximation to the self-energy in which the QP couple only to spin fluctuations. A careful inspection of the different contributions to the RSO self energy shows that the HE kink and the waterfall structure is due to the coupling to damped high energy spin excitations.

{\em{Formalism.}} To study the Hubbard Hamiltonian
we employ the dynamical cluster approximation (DCA)\cite{hettler:dca,maier:rev}, a cluster dynamical mean-field theory which maps the original lattice model onto a periodic cluster of size $N_c=L_c^2$ embedded in a self-consistent host. Correlations up to a range $L_c$ are treated explicitly, while those at longer length scales are described at the mean-field level.  With increasing cluster size, the DCA systematically interpolates between the single-site dynamical mean field (DMFT)\cite{DMFT} and the exact result.  Cluster dynamical mean field calculations of the Hubbard model are found to reproduce many of the features of the cuprates, including a Mott gap and strong AF correlations, d-wave superconductivity and pseudogap behavior~\cite{maier:rev}.  We solve the cluster problem using a quantum Monte Carlo (QMC) algorithm~\cite{jarrell:dca} and employ the maximum entropy method~\cite{jarrell:mem} to calculate the real frequency spectra. The results presented here are obtained from calculations on $N_c=16$ and $N_c=24$ site clusters for $U=8t$.

\begin{figure}[t]
\begin{center}
\includegraphics*[width=3.3in]{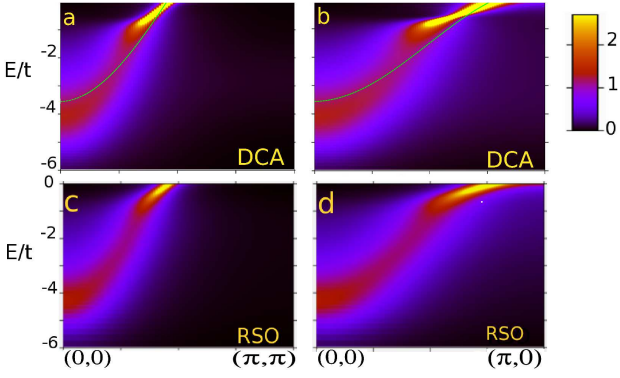}
\caption{(color) Intensity map of the spectral weight $A(k,\omega)$ for $T=0.14t$,   $U=8t$ and $n=0.80$. In (a) and (c), k runs along a nodal $(0,0)-(\pi,\pi)$ cut   and for (b) and (d) k runs along $(0,0)-(\pi,0)$. The DCA results shown in (a)   and (b) were obtained using a 16 site cluster ($N_c=16$) and the RSO results   shown in (c) and (d) were obtained using a self-energy given by Eq.~(1) with   $\bar{U}=0.4U$.  A kink followed by a broad waterfall feature is noticed below   $E_{kink} \approx t$. The thin line in (a) and (b) indicates the bare dispersion   $E(k)$.}
\label{fig:FC}
\end{center}
\end{figure}

{\em{Results.}} The single particle spectral weight of the one-band Hubbard model $A(k,\omega)=-\frac{1}{\pi}ImG(k,\omega)$ reveals a high energy kink in the QP dispersion.  This feature is present for a large range of doping values, from the undoped system up to $\approx 30\%$ doping and along different cuts of the Fermi surface (FS).  In Fig.~\ref{fig:FC} -a and -b we show an intensity map of $A(k,\omega)$ along the diagonal ($(0,0)-(\pi,\pi)$) and center to zone edge ($(0,0)-(\pi,0)$) directions at $20\%$ doping. In both cases an intense QP peak which cuts the FS can be noticed at small energies above $E_{kink} \approx t$. At higher binding energies the dispersion becomes very steep, the peak broadens and decreases in intensity. $E_{kink}$ is weakly decreasing with doping and is weakly dependent on the cut across the FS (not shown). These results are in good agreement with recent experimental findings\cite{j_graf_06a,w_meevasana,t_valla_06,j_chang_06,xie_06,pan_06}.  We find the kink position along the diagonal direction to be at a momentum larger than $(\pi/4,\pi/4)$.  The next-nearest and next-next-nearest neighbor hoppings $t'$ and $t''$, respectively, can however modify the position of the HE kink in the BZ (not shown).  This indicates that $(\pi/4,\pi/4)$ has no particular relevance for the locus of HE kink in the momentum space contrary to some previous suggestions~\cite{j_graf_06a,wang_06,pan_06}. This conclusion is consistent with experimental results reported for $LBCO$~\cite{t_valla_06}.

\begin{figure}[t]
\begin{center}
\includegraphics*[width=3.3in]{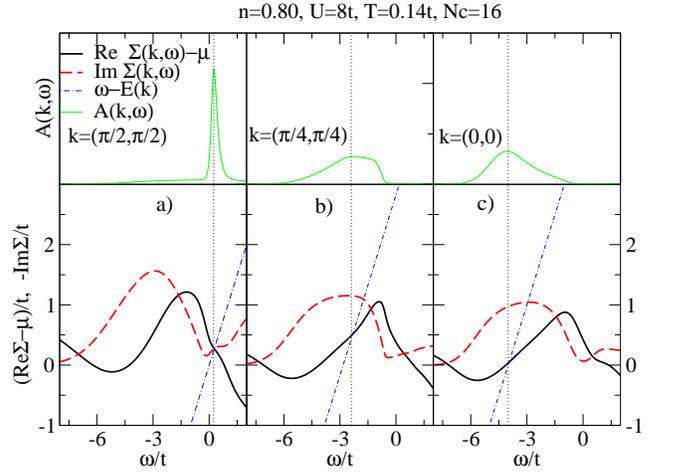}
\caption{(color online) Real part $Re\Sigma(\k,\omega)-\mu$ (thick full) and imaginary part $   -Im \Sigma(\k,\omega)$ (dashed) of the self-energy at a) $\k=(\pi/2,\pi/2)$, b)   $\k=(\pi/4,\pi/4)$ and c) $\k=(0,0)$.  The peak in $A(k,\omega)$ (thin line)   corresponds to the intersection of $Re\Sigma(\k,\omega)-\mu$ with $\omega-E(\k)$  (dashed-dotted). The QP is well defined down to $-\omega =E_{kink}$ where   $Im\Sigma(\k,\omega)$ is small. At larger binding energies a waterfall structure   develops.}
\label{fig:self}
\end{center}
\end{figure}

The HE kink can be inferred from the frequency dependence of the self-energy $\Sigma(\k,\omega)$. In Fig.~\ref{fig:self} we show $\Sigma(\k,\omega)$ at three different $\k$ points along the diagonal direction.  At the kink energy the $\k$ dependence of the self-energy is weak.  Starting from the Fermi energy and increasing $-\omega$, $Re \Sigma(\k,\omega)$ has a negative slope characteristic of a QP with an enhanced effective mass.  The QP is positioned at the intersection of $\omega-E(\k)+\mu$ with $Re \Sigma(\k,\omega)$ and is sharp (see Fig.~\ref{fig:self}-a), a consequence of a small $Im \Sigma(\k,\omega)$.  Here $E(\k)=-2t(\cos k_x +\cos k_y )$ is the bare dispersion.  This QP feature persists down to an energy $-\omega =E_{kink}$ where $Re \Sigma(\k,\omega)$ shows a maximum. At larger binding energies $Re \Sigma(\k,\omega)$ has a positive slope which results in the steep dispersion characterizing the waterfall region seen in ARPES. The slope increases with a finite $t'$ resulting in a steeper waterfall dispersion (not shown).  However in the waterfall region, $Im \Sigma(\k,\omega)$ is large yielding a broad and asymmetric feature in $A(k,\omega)$ (see Fig.~\ref{fig:self} -b $\&$ c), with the  maximum still at the 
intersection of $\omega-E(\k)+\mu$ with $Re \Sigma(\k,\omega)$.
The region where $Re \Sigma(\k,\omega)$ has a positive slope spans a large energy range, between $-t$ and $-6t$, thus characterizing the spectrum down to the $\Gamma$ point at the bottom of the band (Fig.~\ref{fig:self} -c).  It is interesting to notice that the asymmetry of the spectral feature below $E_{kink}$ in Fig.~\ref{fig:self} -b and -c reveals the existence of two maxima in $A(k,\omega)$ pushed together. In fact these two maxima are much better separated if a finite $t'$ is considered, one with a steep dispersion and the other with a strongly renormalized one. Similar behavior has been seen in experiment (see Fig.4 in\cite{j_graf_06a}).  We will present results for the Hubbard model with higher order hoppings elsewhere.

The DCA results for the HE kink are different from the results of other approaches such as the four-pole approximation\cite{grober_00,odashima_05} which considers the scattering of the QP in the lower Hubbard band on spin excitations, or the DMFT\cite{K. Byczuk_06}. These studies find a $Re \Sigma(\k,\omega)$ in the waterfall region
with a slope much larger than one, and hence the corresponding $A(k,\omega)$ displays a gap between two distinct bands. Here, 
we find $0 < \partial Re\Sigma(\omega)/\partial \omega < 1$ near the kink, resulting in a dispersive waterfall feature in $A(k,\omega)$.

Since the spin fluctuations are known to be strong in the cuprates, a reasonable assumption for the origin of the HE kink is the scattering of QP with spin excitations\cite{grober_00,odashima_05,ronning_05,manousakis_06,t_valla_06}.  In order to understand the origin of the HE kink in our results, we therefore test how well a simple renormalized second order (RSO) approximation to the self-energy given by
\begin{equation}
\label{eq:rso_self}
\Sigma^{RSO}(\K,i\omega)= \frac{3}{2}\bar{U}^2\sum_Q \sum_{\nu} 
G_c(\K-\Q,i\omega-i\nu) \chi_c (Q,i\nu)~\,,
\end{equation} 
analytically continued to real frequencies, can describe the ``exact'' DCA self-energy. In Eq.~(\ref{eq:rso_self}), $G_c(\K,i\omega)$ and $\chi_c(\Q,i\nu)$ are the interacting cluster DCA Green's function and spin susceptibility respectively and $\bar{U}$ is a renormalized interaction vertex\cite{m_norman_87,a_kampf_90}. $\K$ and $\Q$ are the cluster momenta\cite{maier:rev}. This approximation neglects the frequency and momentum corrections to the interaction vertex, and the contributions from the charge and pairing channels which we find considerably smaller than the contribution from the spin channel.
By comparing the DCA results for $A(k,\omega)$ in Fig.~\ref{fig:FC} -a and b with the corresponding spectra calculated with the RSO approximation shown in Fig.~\ref{fig:FC} -c and -d, respectively, one can see that the HE kink is well captured by the spin RSO approximation. To obtain this agreement we have set $\bar{U}=0.4 U$. We also find good agreement between the DCA and RSO 
results for $A(k,\omega)$ at $5\%$ doping  with $\bar{U}=0.3 U$ (not shown).

\begin{figure}
\begin{center}
\includegraphics*[width=3.3in]{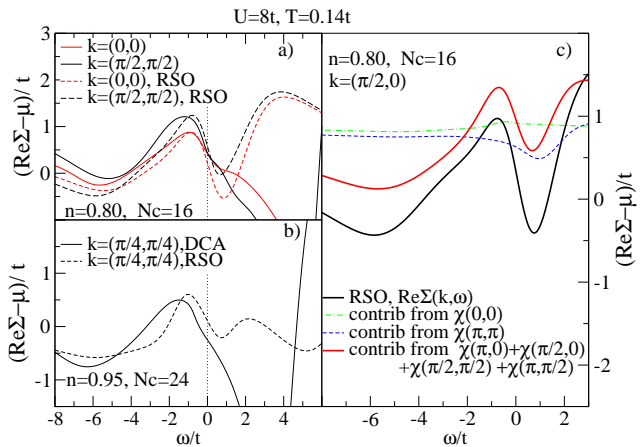}
\caption{(color) a) $Re\Sigma(\k,\omega)-\mu$  at $\K=(\pi/2,\pi/2)$ (black) 
and $\K=(0,0)$ (red) at $20\%$ doping  a)  and 
at $\K=(\pi/4,\pi/4)$ and  $5\%$ doping b).
The full (dashed) lines are the DCA (RSO, Eq.\ref{eq:rso_self}) results. 
c)  Contribution to $Re\Sigma^{RSO}(\K,\omega)$ (black) with $\K=(\pi/2,0)$
from spin excitations with different momentum $\Q$.
The high energy spin excitations 
(red) yield a maximum at  $\omega=-E_{kink}$. 
The low energy spin excitations with $\Q=(0,0)$ (green) and $\Q=(\pi,\pi)$ (blue) 
have a negligible contribution to the HE kink.}
\label{fig:RSO}
\end{center}
\end{figure}

The similarity between the DCA and RSO spectra can be deduced from the corresponding self-energies.  $Re\Sigma^{RSO}(\K,\omega)$ is shown in Fig.~\ref{fig:RSO} -a with dashed lines at $\K=(\pi/2,\pi/2)$ and $\K=(0,0)$ at $20\%$ doping.  Like the DCA self-energy (full lines) $Re\Sigma^{RSO}(\K,\omega)$ shows a maximum at $\omega=-E_{kink}$ which causes the kink seen in the QP dispersion.  The DCA and RSO self-energies agree well over the energy range relevant for the HE kink, especially in the optimally doped and overdoped regions ($15\%-30\%$ doping).  The agreement is still good at small doping as can be seen from Fig.~\ref{fig:RSO} -b, where the $5\%$ doping case at $\K=(\pi/4,\pi/4)$ is shown.  However at small doping the RSO self-energy gives a smaller $E_{kink}$ and a steeper waterfall dispersion, presumably due to the neglect of strong coupling effects which become important in this region.  At positive $\omega$ of order of several $t$ the RSO self-energy differs from the DCA one, resulting in an underestimation of the Mott gap.

A careful analysis of the different $\Q$ contributing to the RSO self-energy in Eq.~(\ref{eq:rso_self}) shows that the HE kink is caused by scattering from high energy spin excitations.  As an example, the red line in Fig.~\ref{fig:RSO} -c is the net contribution to the real part of the self-energy at $\K=(\pi/2,0)$ from $\chi_c(\Q,\nu)$ with $\Q=(\pi,\pi/2)$, $\Q=(0,\pi/2)$, $\Q=(\pi/2,\pi/2)$ and $\Q=(0,\pi)$\cite{DCA_chi}.  It displays a maximum at $\omega=-E_{kink}$, the typical energy associated with spin excitations at the magnetic zone boundary (see next paragraph), thus yielding the HE kink.  The low energy spin excitations at the zone center (green) and zone corner (blue) do not contribute to the maximum in $Re\Sigma^{RSO}(\K,\omega)$ and therefore are not important for the HE kink.  The analysis of $Re\Sigma^{RSO}(\K,\omega)$ at other $\K$ values  leads to the same conclusions.

\begin{figure}[t]
\begin{center}
\includegraphics*[width=3.3in]{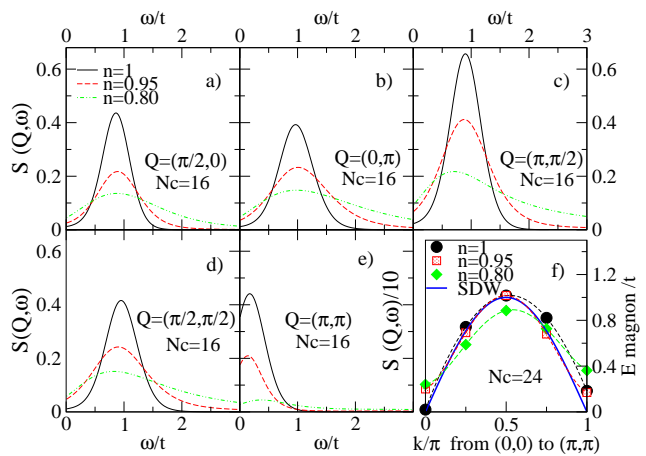}
\caption{(color online) a), b), c), d) $\&$ e) Doping dependence of $S(\Q,\omega)$ for   different $\Q$ in the BZ. 
High energy spin excitations persist when the doping is increased, displaying a maximum at $\omega \approx 2J$ in $S(\Q,\omega)$.
In e) $S((\pi.\pi),\omega)$ is scaled with a factor of $0.1$.  f) Dispersion of   the magnon peak along the diagonal 
direction for different dopings. Here $U=8t$ and $T=0.14t$.}
\label{fig:magnon}
\end{center}
\end{figure}

Short range spin excitations with a characteristic energy $2J\approx 8t^2/U$ persist up to relative large doping.  The magnetic structure factor $S(\Q,\omega)$ for different values of the doping is shown in Fig.~\ref{fig:magnon} -a, -b, -c, -d, $\&$ -e at different cluster $\Q$ in the BZ. In the undoped system $S(\Q,\omega)$ shows sharp magnon peaks with an energy predicted in agreement with the spin-density wave (SDW) approximation \cite{manousakis_rmp}, as can be seen in Fig.~\ref{fig:magnon} -f where the magnon dispersion along the diagonal direction is shown\cite{magnon_dca}. 
With increasing doping $S(\Q,\omega)$ broadens and, in the region of the BZ corresponding to high energy spin excitations,
still retains a well defined maximum  at an energy of order of $\approx 2J$, as shown in Figs.~\ref{fig:magnon} -a through -d.
 In this region of the BZ the total weight of $S(\Q,\omega)$ does not change much with increasing doping but a significant transfer of 
weight to higher energies can be noticed.  For instance the peaks in $S(\Q,\omega)$ at $\Q=(0,\pi)$ and $\Q=(0,\pi/2)$ are positioned at $\approx 2J$ for $0\%$, $5\%$ and $20\%$ doping. The magnon peaks at $\Q=(\pi/2,\pi)$ and $\Q=(\pi/2,\pi/2)$ soften a little with doping, presumably causing the small decrease in $E_{kink}$ with doping.  However, we find that the low energy spin excitations around $\Q=(\pi,\pi)$ are more strongly affected by doping as may be seen in Fig~\ref{fig:magnon} -e. The total spectral weight is strongly reduced with doping. At $20\%$ doping $S((\pi,\pi),\omega)$ shows a broad peak with a maximum at an energy $\approx J$.

{\em{Discussions.}} These results suggest that the HE kink is caused by  coupling to damped high-energy spin fluctuations. The dispersive spectral feature in the waterfall region is a consequence of $Re\Sigma$ with $0<\partial Re\Sigma(\omega)/\partial \omega < 1$.  This requires scattering on damped excitations with an energetically broad spectrum.  As seen in Fig.\ref{fig:self}, the scattering rate $-Im\Sigma$ initially increases with $-\omega$.  However, at higher binding energies $Im\Sigma$ passes through a maximum and decreases as the phase space for the final scattering states decreases.  Within the RSO approximation this reflects the convolution of the spin-fluctuation particle-hole continuum with $A(k,\omega)$ and leads, in the present case, to a peak in $-Im\Sigma$ for $\omega\approx-3t$. The structure in the $Re\Sigma$ follows from the Kramers-Kronig relation and can be understood a result of energy level repulsion. At small values of $-\omega$ the majority of the states in the single-particle-spin-fluctuation convolution have energies larger than $-\omega$ and give rise to the usual QP mass enhancement.  However, at larger values of $-\omega$, the dominant contribution from these intermediate states comes from states with lower energies leading to the decrease in $Re\Sigma$ and driving the dispersion of the spectral feature at high binding energies below the bare dispersion (thin line in Fig.\ref{fig:FC}).

While the main features of HE kink and waterfall can be captured with a single-band model with $U \approx W$,  comparison with experiment requires realistic values for the Hamiltonian parameters. We already mentioned that a next-nearest neighbor hopping $t'$ makes the waterfall dispersion steeper, presumably due to sharper spin excitations\cite{macridin_arpes}. A $t''$ hopping moves $\k_f$ and the locus of HE kink on the diagonal direction in BZ towards the $\Gamma$-point.  We also find that $E_{kink}$ decreases with increasing U presumably due to the reduction of effective $J$ (not shown).  At high energy, the experimental ARPES in cuprates show oxygen valence states in the proximity to the $\Gamma$ point\cite{j_graf_06a,w_meevasana}, which obviously are not captured with a single-band Hubbard model.  Moreover other states missing in the single band model, such as the $d_{3z^2-r^2}$ states, should also be considered when analyzing the experimental ARPES spectra below $0.5~eV$, as multi-orbital calculations for cuprates indicate\cite{eskes}.

The simple renormalized second order ansatz, Eq.~\ref{eq:rso_self}, seems to provide a good description of the single-particle ARPES spectra of the Hubbard model with parameters relevant for the cuprates outside the pseudogap regime.  This suggests that this ansatz could be used to analyze experiments where $\chi(q,\omega)$ is measured by neutron scattering, and used to construct the ARPES spectra.  $\bar U$ could be fixed by fitting the RSO spectra to the high energy kink. Consistency between the measured and constructed spectra would strongly suggest that the HE kink in the experimental ARPES spectra may be described with a single-band model and is due to the coupling to spin fluctuations.  As discussed elsewhere\cite{th_maier_06}, a similar RSO result also provides an accurate description of the pairing interaction of the Hubbard model in the regime relevant for the cuprates. Thus the neutron spectra, together with the $\bar U$ obtained from the fit above, could be used in a simple RSO calculation of the superconducting phase diagram.

{\em{Conclusions.}}  By employing DCA calculations we show that the single-band Hubbard model captures the HE kink structure seen in the cuprates.  The kink occurs as a crossover from a well defined QP peak to a waterfall structure characterized by a broad and asymmetric feature with steep dispersion.  The structure of the HE kink is well captured by a simple renormalized second order self-energy which couples the quasiparticle to spin fluctuations. A careful decomposition of the contributions to the RSO self-energy indicate that the HE kink and the waterfall structure in the spectrum of the Hubbard model is due to the damped high energy spin fluctuation continuum.

\acknowledgments We thank T.\ Devereaux, A.\ Lanzara, W.\ Meevasana, B.\ Moritz, G.\ A.\ Sawatzky and F.\ C.\ Zhang for useful discussions. This research was supported by NSF DMR-0312680 and CMSN DOE DE-FG02-04ER46129.  Supercomputer was provided by NSF SCI-9619020 through resources provided by the San Diego Supercomputer Center and the National Center for Computational Sciences at Oak Ridge National Laboratory, supported by the Office of Science of the U.S. Department of Energy under Contract No. DE-AC05-00OR22725. TM and DJS acknowledges the Center for Nanophase Materials Sciences, which is sponsored by the Division of Scientific User Facilities, U.S. Department of Energy.


\begin{thebibliography}{99}
\bibitem{DCA_chi} In  DCA $\chi_c(\Q, \nu)$ is the average of $\chi(\q, \nu)$
over the momenta in the coarse-grained cell  centered on $\Q$\cite{maier:rev}.
\bibitem{magnon_dca} Due to the cutoff of the AF correlations outside the cluster
the  energy of the magnon peak at $\Q=(\pi,\pi)$ at small doping is finite 
and goes to zero with increasing the cluster size.
\end{thebibliography}
\end{document}